\documentclass[runningheads]{cl2emult}

\usepackage{makeidx}  
\usepackage{graphicx} 
\usepackage{subeqnar} 
\usepackage{multicol} 
\usepackage{cropmark} 
\usepackage{math}     
\makeindex            

\begin{document}

\title*{{\it False} EUR Exchange Rates vs. $DKK$, $CHF$, $JPY$ and
$USD$.\protect\newline What is a strong currency?} \toctitle{{\it False} EUR
Exchange Rates vs. $DKK$, $CHF$, $JPY$ and $USD$. \protect\newline What is a
strong currency?} \titlerunning{{\it False} EUR Exchange Rates}

\author{Kristinka Ivanova\inst{1} \and Marcel Ausloos\inst{2}}

\authorrunning{K. Ivanova and M. Ausloos}

\institute{Pennsylvania State University, University Park PA 16802, USA \and
GRASP, B5, University of Li$\grave e$ge, B-4000 Li$\grave e$ge, Belgium}

\maketitle

\begin{abstract} The Euro ($EUR$) has been a currency introduced by 
the European
Community on Jan. 01, 1999. This implies eleven countries of the European Union
which have been found to meet the five requirements of the Maastricht 
convergence
criteria. In order to test $EUR$ behavior and understand various features, we
have extrapolated the $EUR$ backwards and therefore have obtained a {\it false
euro} ($FEUR$) dating back to 1993. We have derived the exchange rates of the
$FEUR$ with respect to several currencies of interest not belonging 
to the $EUR$,
i.e., Danish Kroner ($DKK$), Swiss Franc ($CHF$), Japanese Yen ($JPY$) and U.S.
Dollar ($USD$). We have first observed the distribution of fluctuations of the
exchange rates. Within the {\it Detrended Fluctuation Analysis} ($DFA$)
statistical method, we have calculated the power law behavior describing the
root-mean-square deviation of these exchange rate fluctuations as a function of
time, displaying in particular the $JPY$ exchange rate case. In order 
to estimate
the role of each currency making the $EUR$ and therefore in view of identifying
whether some of them mostly influences its behavior, we have compared the
time-dependent exponent of the exchange rate fluctuations for $EUR$ 
with that for
the currencies that form the $EUR$. We have found that the German Mark ($DEM$)
has been leading the fluctuations of $EUR/JPY$ exchange rates, and Portuguese
Escudo ($PTE$) is the farthest away currency from this point of view.

\end{abstract}

\section{Introduction}

The Euro ($EUR$) is a {\it bona fide} currency introduced by the 
European Community on Jan. 01,
1999 \cite{quoteEUR} in contrast to the $XEU$ which was a theoretical 
''basket'' of currencies. The $EUR$ is superseding national 
currencies in eleven countries of the European Union which have
been found to meet the five requirements of the Maastricht convergence criteria
\cite{quoteEUR}: price stability, fiscal prudence, successful European monetary
system membership, and interest-rates convergence in particular. In 
order to test
$EUR$ behavior and understand various features, we have extrapolated the $EUR$
backwards and therefore have obtained a {\it false euro} ($FEUR$) 
dating back to
1993. We have reconstructed the exchange rates of the $FEUR$ with respect to
several currencies of interest not belonging to the $EUR$, i.e., Danish Kroner
($DKK$), Swiss Franc ($CHF$), Japanese Yen ($JPY$) and U.S. Dollar ($USD$). The
$DKK$ is a currency for a country belonging to the European Community 
and outside
the $EUR$ system. The $CHF$ is a European currency for a country NOT 
belonging to
the European system. The $JPY$ and $USD$ are both major currencies outside
Europe.

The irrevocable conversion rates of the participating countries have been fixed
by political agreement based on various considerations and the bilateral market
rates of December 31, 1998 \cite{quoteEUR,rateEUR,Ref1EUR}. Using 
these rates one  {\it false}
Euro ($FEUR$) can be represented as an unweighted sum of the eleven currencies
$C_i$, $i= 1, 11$:

\begin{equation} 1 EUR = \sum_{i=1}^{11} \frac{\gamma_i}{11} \, C_i
\end{equation}

\noindent where $\gamma_i$ are the conversion rates and $C_i$ denote the
respective currencies, i.e. Austrian Schilling ($ATS$), Belgian Franc ($BEF$),
Finnish Markka ($FIM$), German Mark ($DEM$), French Franc ($FRF$), Irish Pound
($IEP$), Italian Lira ($ITL$), Luxembourg Franc ($LUF$), Dutch Guilder ($NLG$),
Portuguese Escudo ($PTE$), Spanish Peseta ($ESP$). In order to study 
correlations
in the $EUR$ exchange rates as of now, the $EUR$ existence can be artificially
extended backward, i.e., before Jan. 01, 1999. This can be done by applying
Eq.(1) to each participating currency for the time interval of the 
exchange rates
which are available before Jan. 01, 99, thereby defining a more or less legal
(but $false$) $FEUR$ before its birth. Nevertheless we drop the letter $F$ in
$FEUR$ thereafter.

We are concerned with the behavior of $EUR$ toward currencies which are outside
the European Union, since nowadays these are the only exchange rates 
of interest.
These are e.g. Danish Kroner ($DKK$), Swiss Franc ($CHF$), Japanese Yen ($JPY$)
and U.S. Dollar ($USD$). Therefore, we construct a data series of $EUR$ toward
e.g. Japanese Yen ($EUR$/$JPY$) following the artificial rule:

\begin{equation} 1 EUR/JPY = \sum_{i=1}^{11} \frac{\gamma_i}{11} \, C_i/JPY
\end{equation}

However, the number of data points of the exchange rates for the 
period starting
Jan. 1, 1993 and ending Dec. 31, 1998 is different for these eleven currencies
toward $DKK$, $CHF$, $JPY$ and $USD$. This is due to different 
national and bank
holidays when the banks are closed and official exchange rates are 
not defined in
some countries. The number $N$ of data points has been equalized as done in
\cite{Ref1EUR}, assuming that the exchange rate does not change if 
there is such
a gap (usually a holiday), such that $N = 1985$, spanning the 
interval time from
January 1, 1993 till October 31, 2000.

The evolution of the $false$ (from Jan. 1, 1993 to Dec. 31, 1998) and the real
(from Jan. 1, 1999 to Jun 30, 2000) $EUR$ with respect to the $DKK$, $CHF$,
$JPY$, and $USD$ are plotted in Fig. 1. While the $EUR/DKK$ exchange 
rate is not
much disturbed by the transition to the real $EUR$, the other currencies, in
particular $USD$ and $JPY$ and a little bit less $CHF$ have been much sensitive
to the transition.

\begin{figure} \centering \includegraphics[width=.7\textwidth]{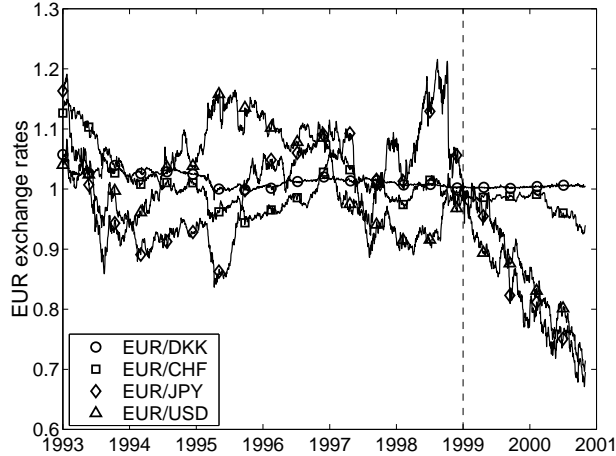}
\caption{Normalized evolution assuming an exchange rate = 1 on Jan. 
01, 1999 for
$EUR/DKK$, $EUR/CHF$, $EUR/JPY$ and $EUR/USD$} \label{eps1} \end{figure}

The $DFA$ technique \cite{DFA} has been often described and is not 
recalled here.
It leads to investigating whether the root-mean-square deviations of the
fluctuations of the investigated signal $y(n)$ has a scaling behavior, e.g. if
the function $<F^2(\tau)>$ scales with time as

\begin{equation} \langle {1 \over \tau } {\sum_{n=k\tau+1}^{(k+1)\tau}
{\left[y(n)- z(n)\right]}^2} \rangle \sim \tau^{2\alpha} \end{equation}

\noindent where $z(n)$ is hereby a linear function fitting at best the data in
the $\tau$ interval which is considered. A value $\alpha = 0.5$ 
corresponds to a
signal mimicking a Brownian motion.

Let it be recalled that in \cite{Ref1EUR} it has been shown that the time scale
invariance for $EUR/CHF$, $EUR/USD$, and $EUR/JPY$ holds from 5 days (one week)
to about 300 days (one year) showing Brownian type of correlations. 
Two different
scaling ranges were found for the $EUR/DKK$; one, from four to 25 
days (5 weeks)
with a non-Brownian $\alpha=0.37 \pm 0.01$, and another, after that 
for up to 300
days (61 weeks) with Brownian-like correlations.

In order to estimate the role of each currency making the $EUR$ and 
therefore in
view of identifying whether some of them mostly influences its 
behavior, we have
first looked at the distribution of exchange rate fluctuations in the 
interesting
time interval defined above. Next the time-dependence of the $\alpha$ exponent
characterizing the scaling law for the exchange rate fluctuation 
correlations for
$EUR$ and that for the 11 currencies that form $EUR$ has been calculated. This
evolution has been averaged (i) over the currencies, (ii) over the 
time interval
considered here. The results are compared here to the behavior of the 
fluctuation
correlations for $EUR/JPY$ and $DEM/JPY$ exchange rates. Let it be pointed out
that we have also tested elsewhere the $DKK$, $CHF$ and $USD$ 
exchange rates with
respect to $EUR$, $DEM$ and the other $EUR$ forming currencies 
\cite{kimaijmpc}.

\section{Distribution of the fluctuations and strong currency}

It is of interest to observe whether the usual statement that the 
$EUR$ {\it is
nothing else than a generalized }$DEM$ holds true. In order to do so we have
first compared the distributions of the exchange rate fluctuations 
for $DEM/DKK$,
$DEM/CHF$, $DEM/JPY$, $DEM/USD$ with the distributions of the fluctuations for
$EUR/DKK$, $EUR/CHF$, $EUR/JPY$, $EUR/USD$ (Fig. 2). From such a comparison we
are led to consider that $DEM$ is dominant in defining the distribution of the
fluctuations of $EUR$ exchange rate with respect to $DKK$, $CHF$, $JPY$ and
$USD$. For all cases the central part of the distributions can be fitted by a
Gaussian distribution while the tails, i.e. the large fluctuations, follow a
power law with a slope equal to $ca.$ 2.9 for $EUR/CHF$ and $DEM/CHF$, 3.2 for
the negative and 4.0 for the positive tail of $EUR/JPY$ and $DEM/JPY$, and 3.2
for the negative and about 4.5 for the positive tail in $EUR/USD$ and 
$DEM/USD$.

It is fair to recall that the volatility of exchange rates follows different
scaling laws depending on the horizon which is considered \cite{friedrich}.
However the correlation coefficient stabilizes at scales one day and higher
\cite{gencay}. Those values might be examined whether they result from the
equivalent of trading momentum and price resistance just like in the model of
\cite{castiglione} for stock price and price fluctuation distribution. The
above asymmetry in the power law for the positive and negative tail 
might result
or not from the limited amount of data.

\begin{figure} \centering \includegraphics[width=.48\textwidth]{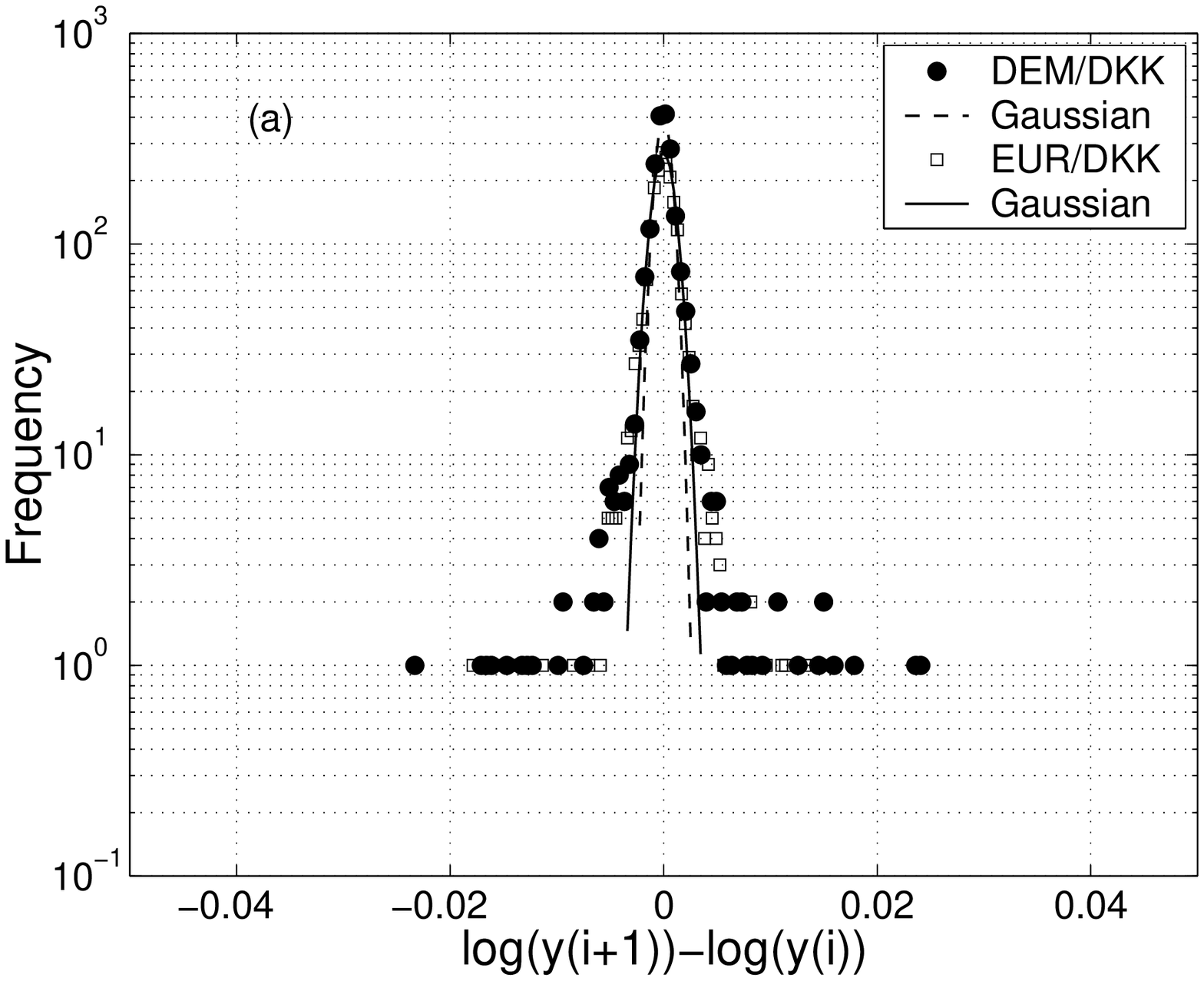}
\hfill \includegraphics[width=.48\textwidth]{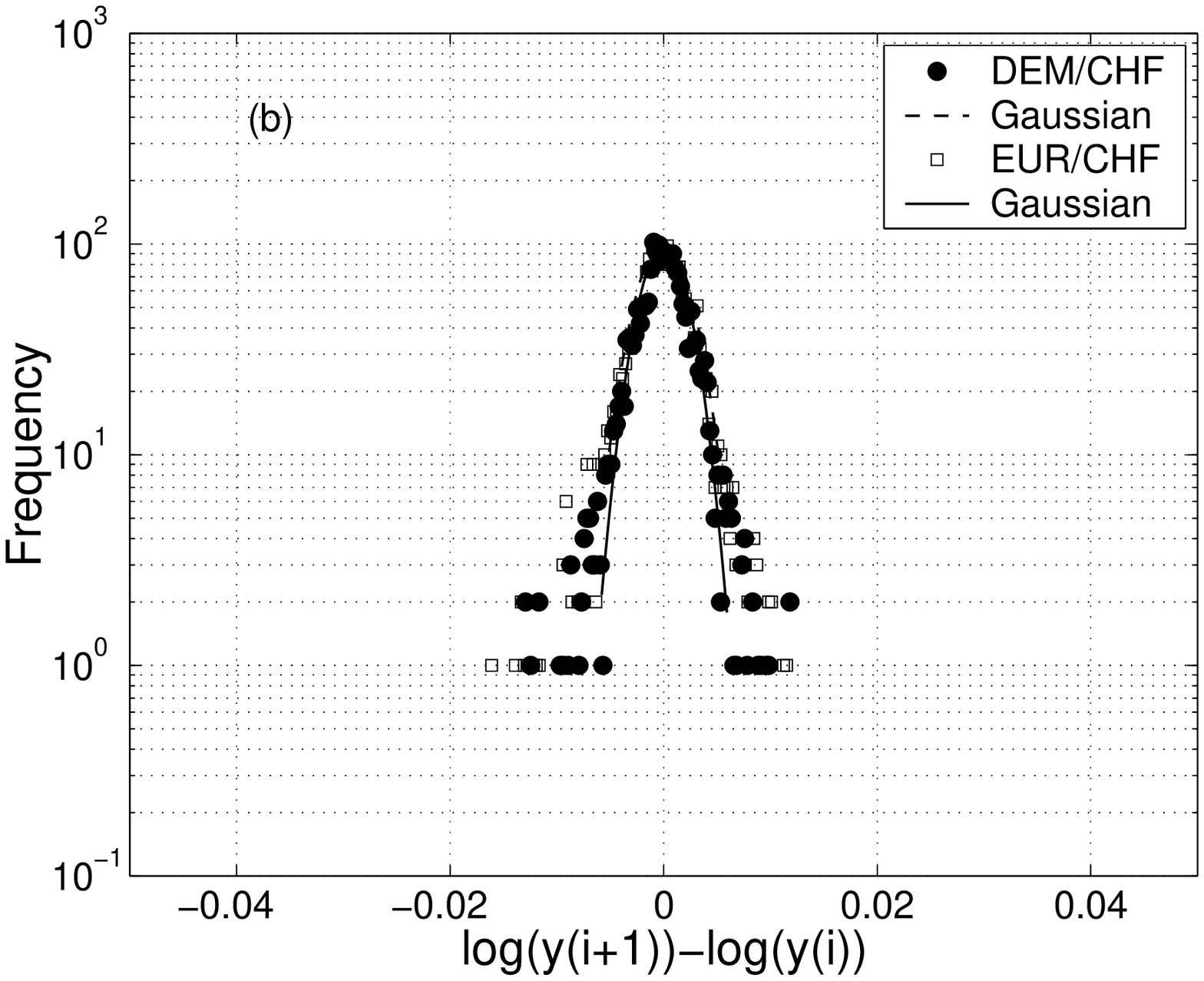} \vfill
\includegraphics[width=.48\textwidth]{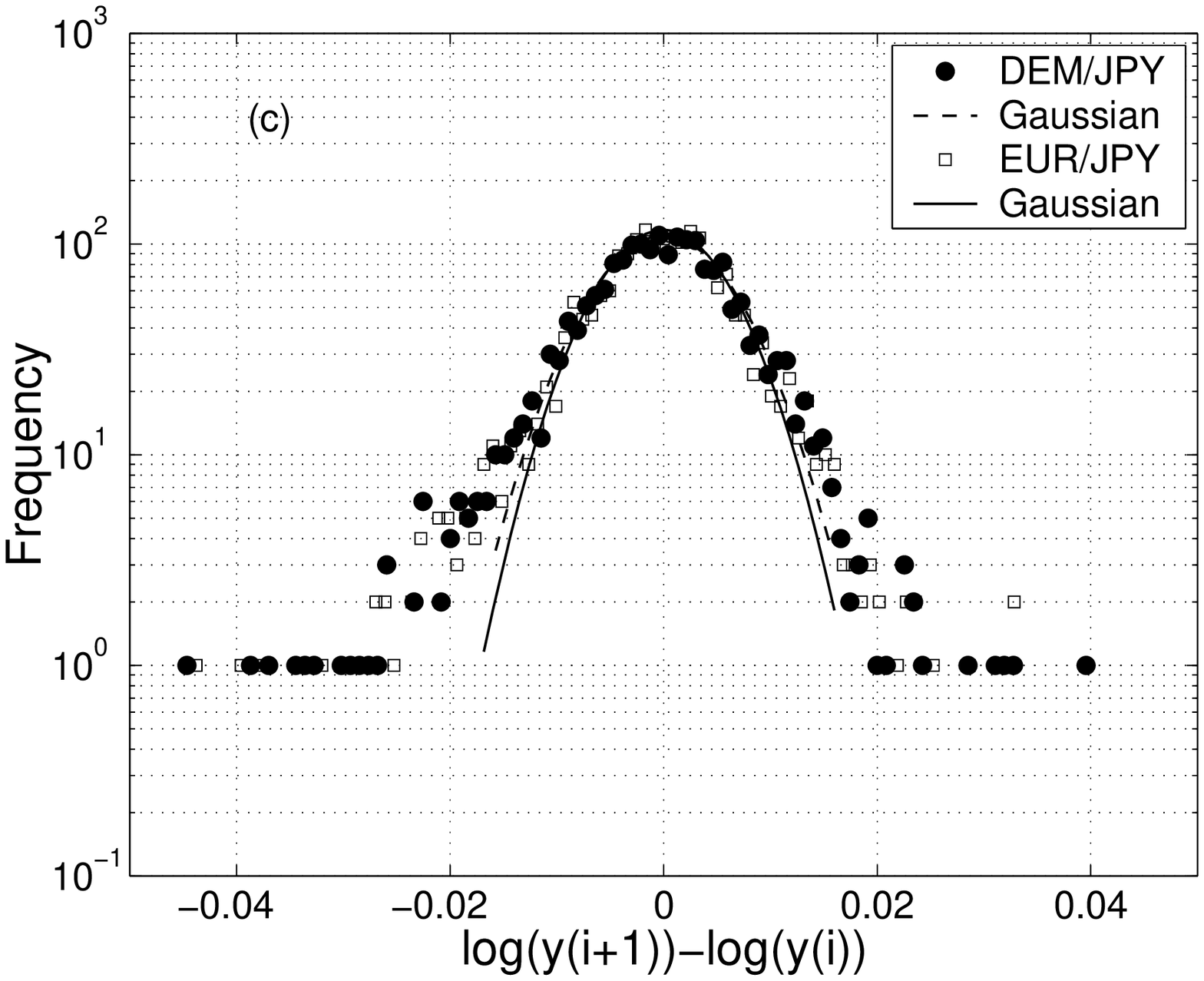} \hfill
\includegraphics[width=.48\textwidth]{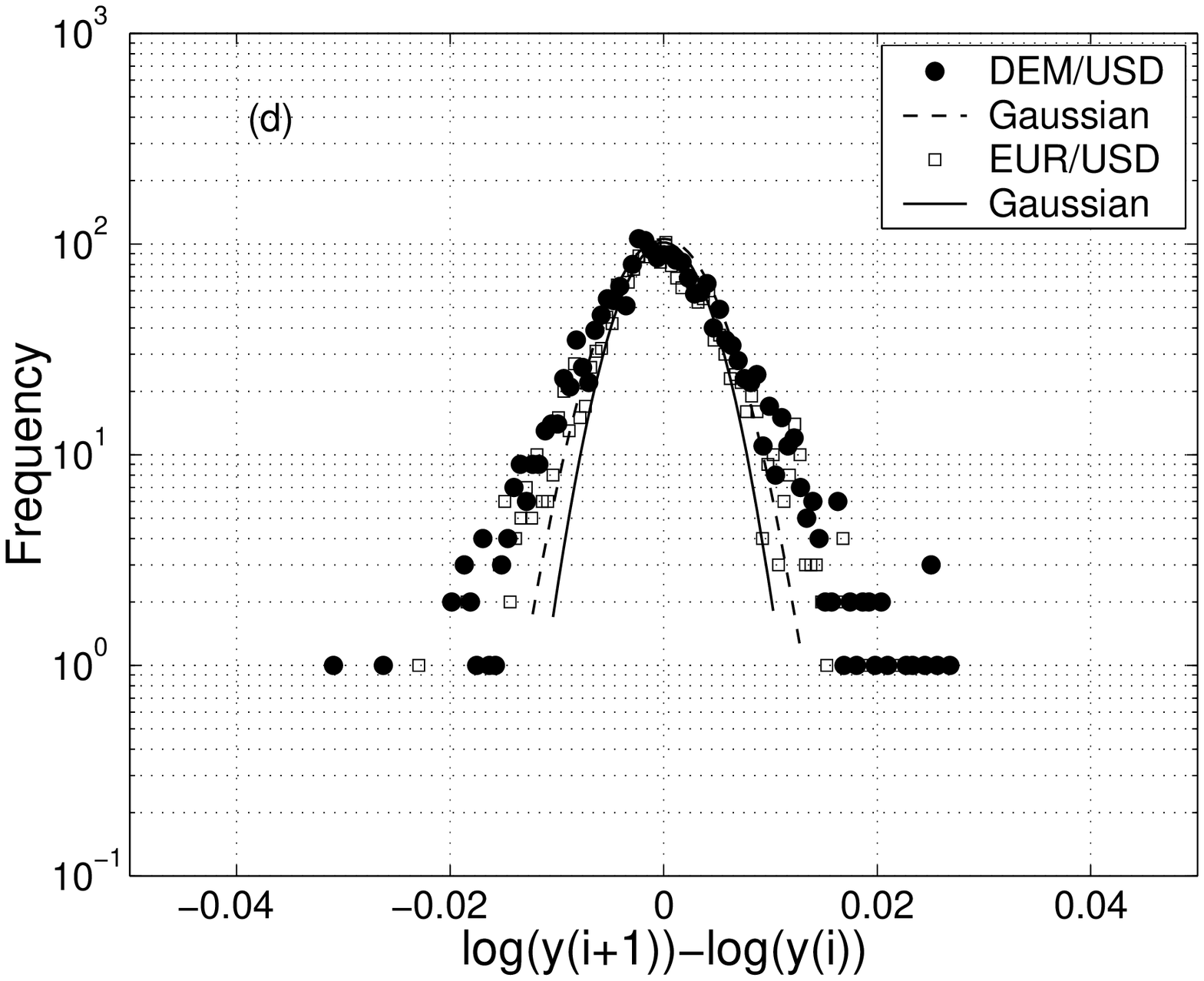} \caption{The distribution of
the fluctuations of {\bf (a)} $DEM/DKK$ and $EUR/DKK$, {\bf (b)} $DEM/CHF$ and
$EUR/CHF$, {\bf (c)} $DEM/JPY$ and $EUR/JPY$, and {\bf (d)} $DEM/USD$ and
$EUR/USD$, as compared to the Gaussian distribution. {\it Dashed lines}
correspond to a Gaussian fit to the $DEM$ exchange rate distributions 
while {\it
solid lines} represent a Gaussian fit to the $EUR$ exchange rate distributions}
\label{eps2} \end{figure}

\section{Correlations of the fluctuations and strong currency}

As done elsewhere \cite{nvma}, in order to probe the existence of {\it locally
correlated} and {\it decorrelated} sequences, we have constructed an 
observation
box, i.e. a 515 days ($\simeq$ 2 years) wide window probe placed at 
the beginning
of the data, calculated $\alpha$ for the data in that box, moved this 
box by one
day toward the right along the signal sequence, calculated $\alpha$ 
in that box,
a.s.o. up to the $N$-th day of the available data. A local, time dependent
$\alpha$ exponent is thus found for the $N-515$ last days. The $JPY$ exchange
rate case results only are illustrated in this report. The time dependent
$\alpha$-exponent for $EUR/JPY$ and that for each of the 11 currency 
(which form
the $EUR$) exchange rates toward $JPY$ have been computed. Together with the
$EUR/JPY$ result, some of the time dependent $\alpha$-exponents, i.e. for
$ATS/JPY$, $DEM/JPY$, $ITL/JPY$, $PTE/JPY$ are shown in Fig. 3 as the most
representative ones of various behaviors. Notice, the similarity between
$EUR/JPY$ and $DEM/JPY$, a maximum in 1998 but a rather flat behavior for
$ATS/JPY$, a very irregular behavior for $ITL/JPY$, and an increase in $\alpha$
around 1998 for $PTE/JPY$. The other cases, i.e.  $BEF/JPY$, $FIM/JPY$,
$FRF/JPY$, $IEP/JPY$, and $NLG/JPY$ are similar to $ATS/JPY$ and 
$DEM/JPY$ cases.
While the differences in the $\alpha$-behavior after Jan. 1, 1999 are almost
undistinguishable, the time dependent $\alpha$-exponents before that 
day exhibit
nevertheless different correlated fluctuations depending on the currency. We
stress that $\alpha$ for $EUR/JPY$ most closely resembles the $DEM/JPY$ before
and after Jan. 01, 99, they are almost identical already since mid 1996.

\begin{figure} \centering \includegraphics[width=.48\textwidth]{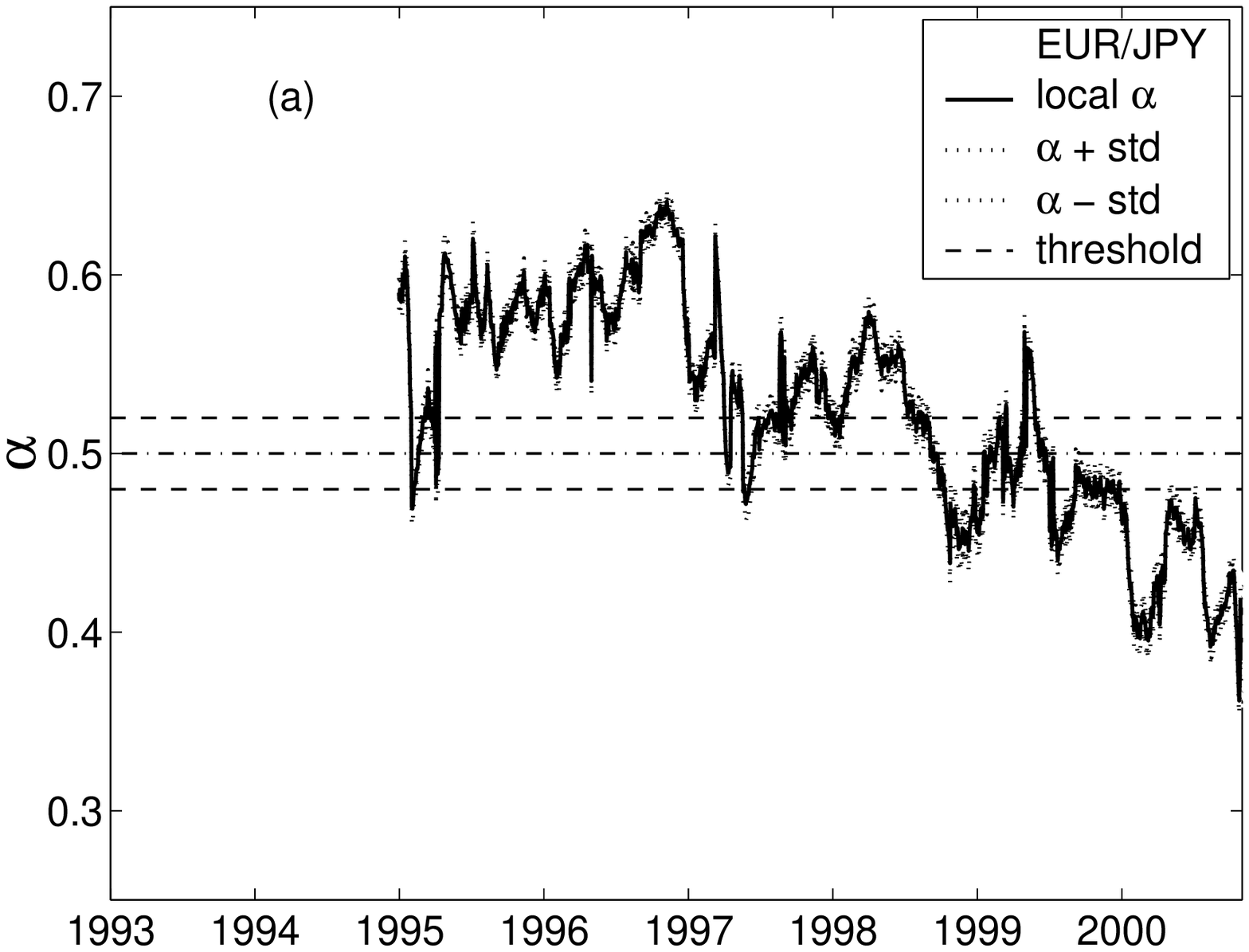}
\vfill \includegraphics[width=.48\textwidth]{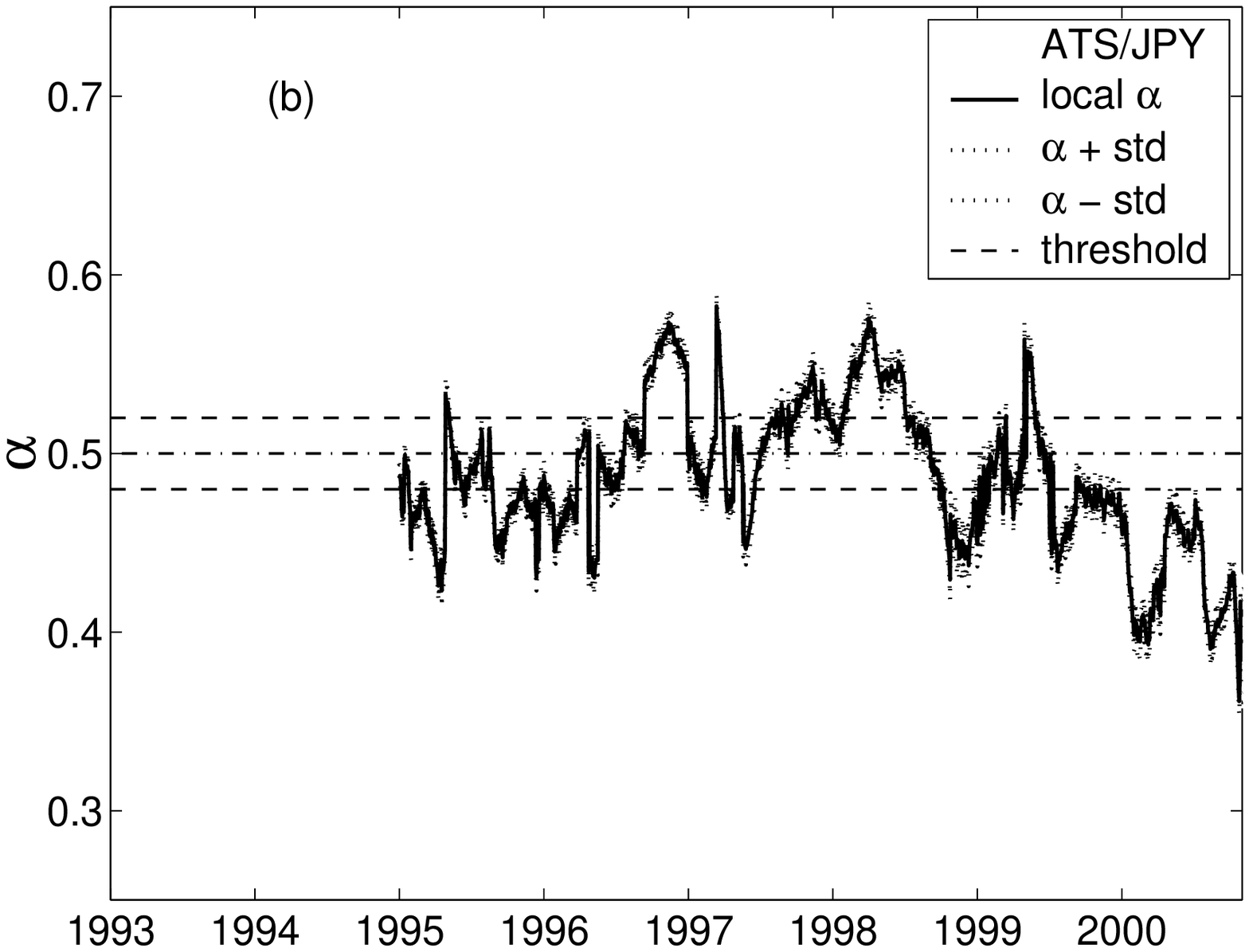} \hfill
\includegraphics[width=.48\textwidth]{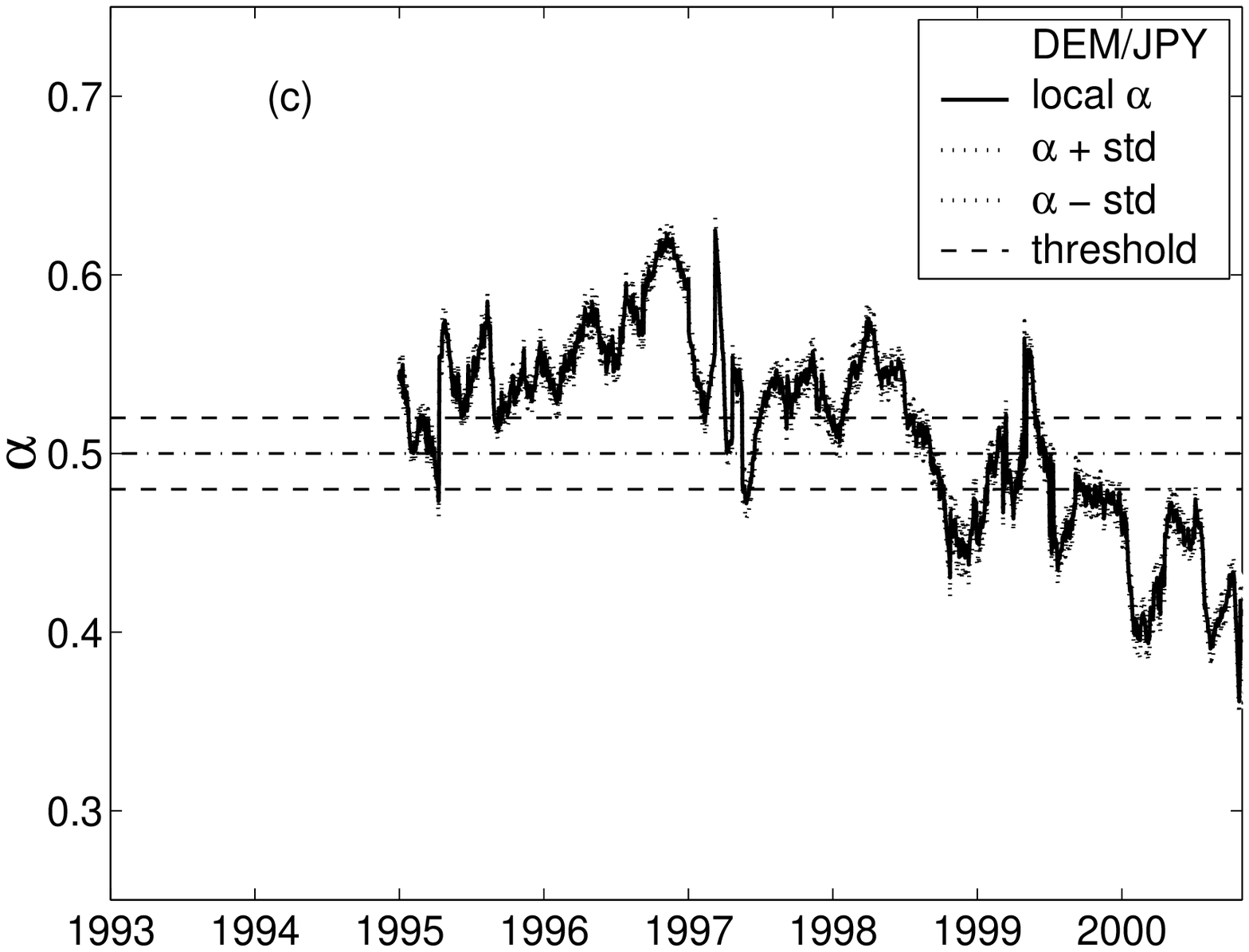} \vfill
\includegraphics[width=.48\textwidth]{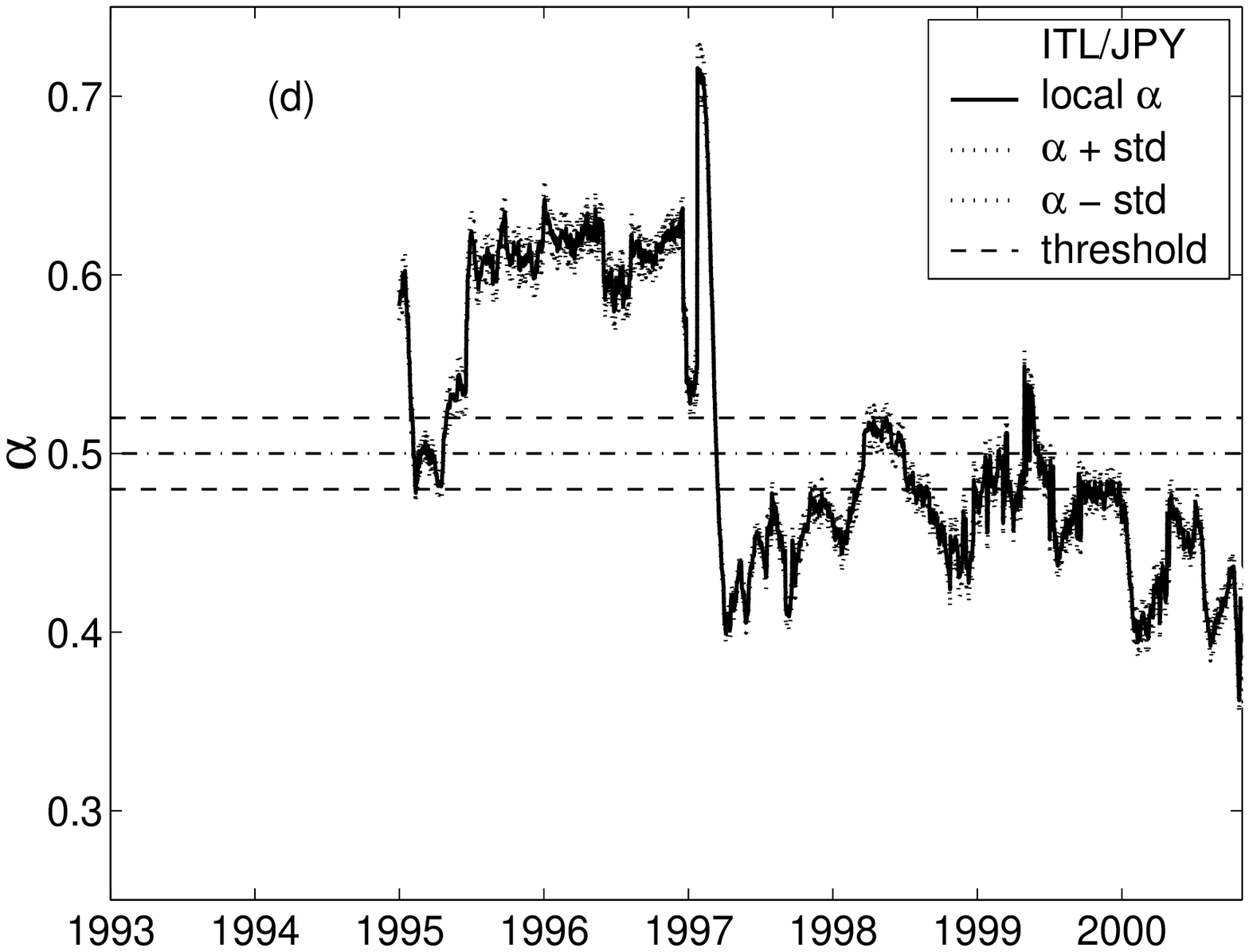} \hfill
\includegraphics[width=.48\textwidth]{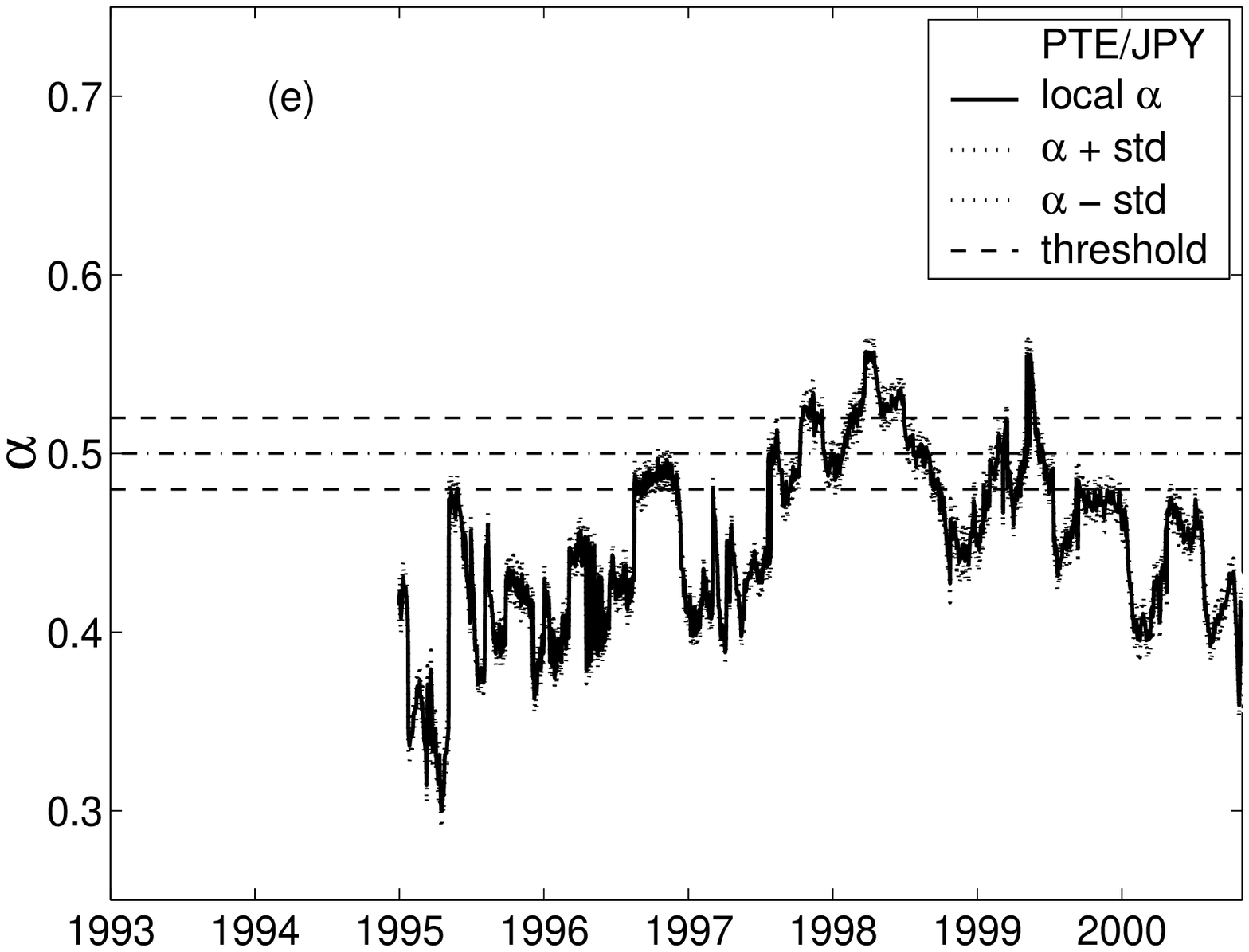} \caption{The evolution of the
local value of $\alpha$ estimated with the $DFA$ method for a window of size 2
years ({\it solid line}) for {\bf (a)} $EUR$, {\bf (b)} $ATS$, {\bf (c)} $DEM$,
{\bf (d)} $ITL$, {\bf (e) $PTE$} with respect to $JPY$. {\it Dotted lines} mark
one standard deviation in the $\alpha$ estimate. {\it Dashed horizontal lines}
indicate a 0.02 threshold relative to the uncorrelated fluctuations 
of a Brownian
motion signal, $\alpha = 0.5$} \label{eps3} \end{figure}

Notice that the $\alpha$-exponent for $EUR/JPY$ and those 
$\alpha$-values of the
other currency exchange rates are not strictly equal to each other even after
Jan. 1, 1999 because the fit window used for calculating $\alpha$ includes days
prior to Jan. 1, 1999. A strict identity should only occur on Jan. 1, 
2001, i.e.
after two years (515 days) from $EUR$ strict birth. Yet, the values are already
very close to each other (less than 1\%) before such a date.

\begin{figure} \centering \includegraphics[width=.36\textwidth]{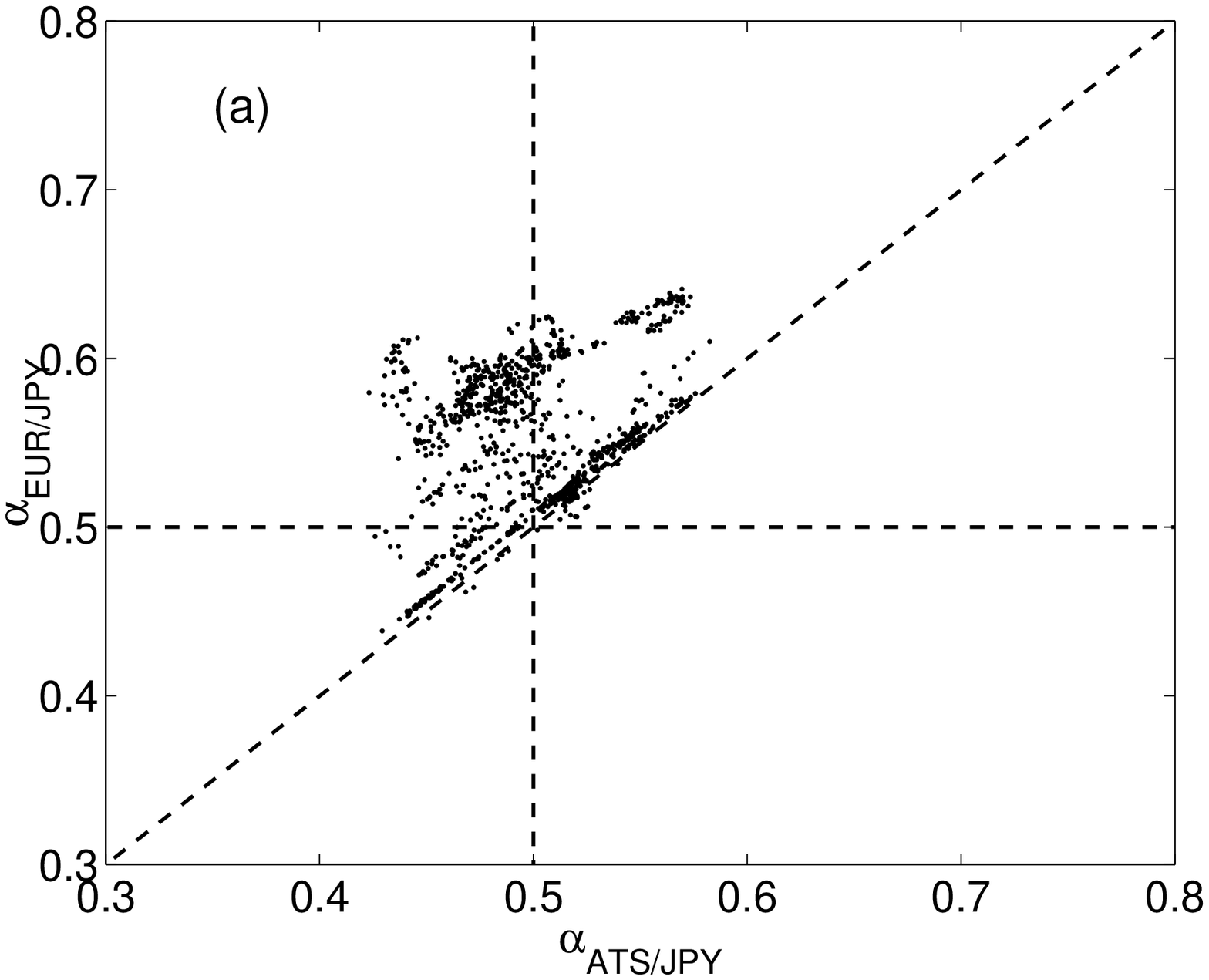}
\hfill \includegraphics[width=.36\textwidth]{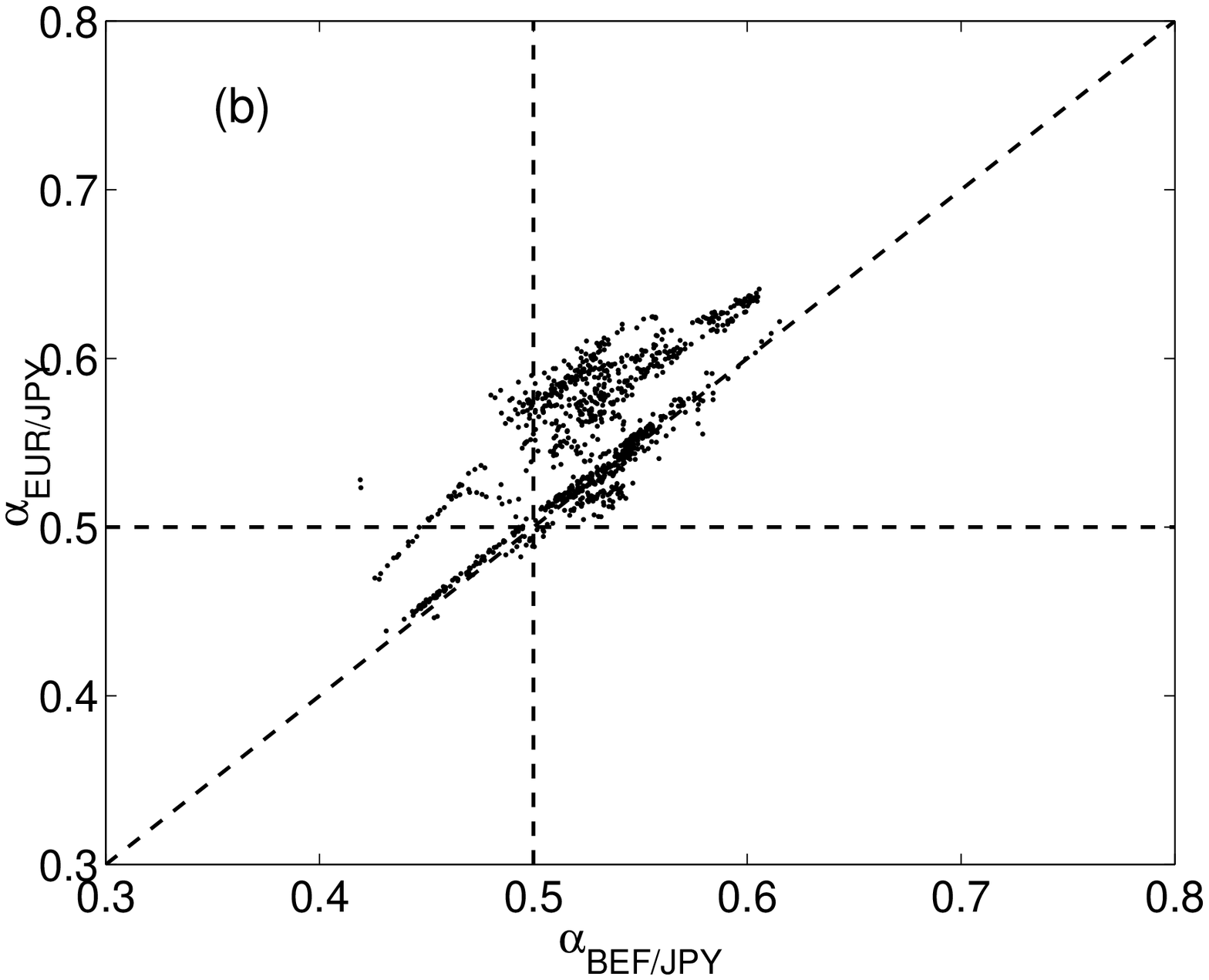} \vfill
\includegraphics[width=.36\textwidth]{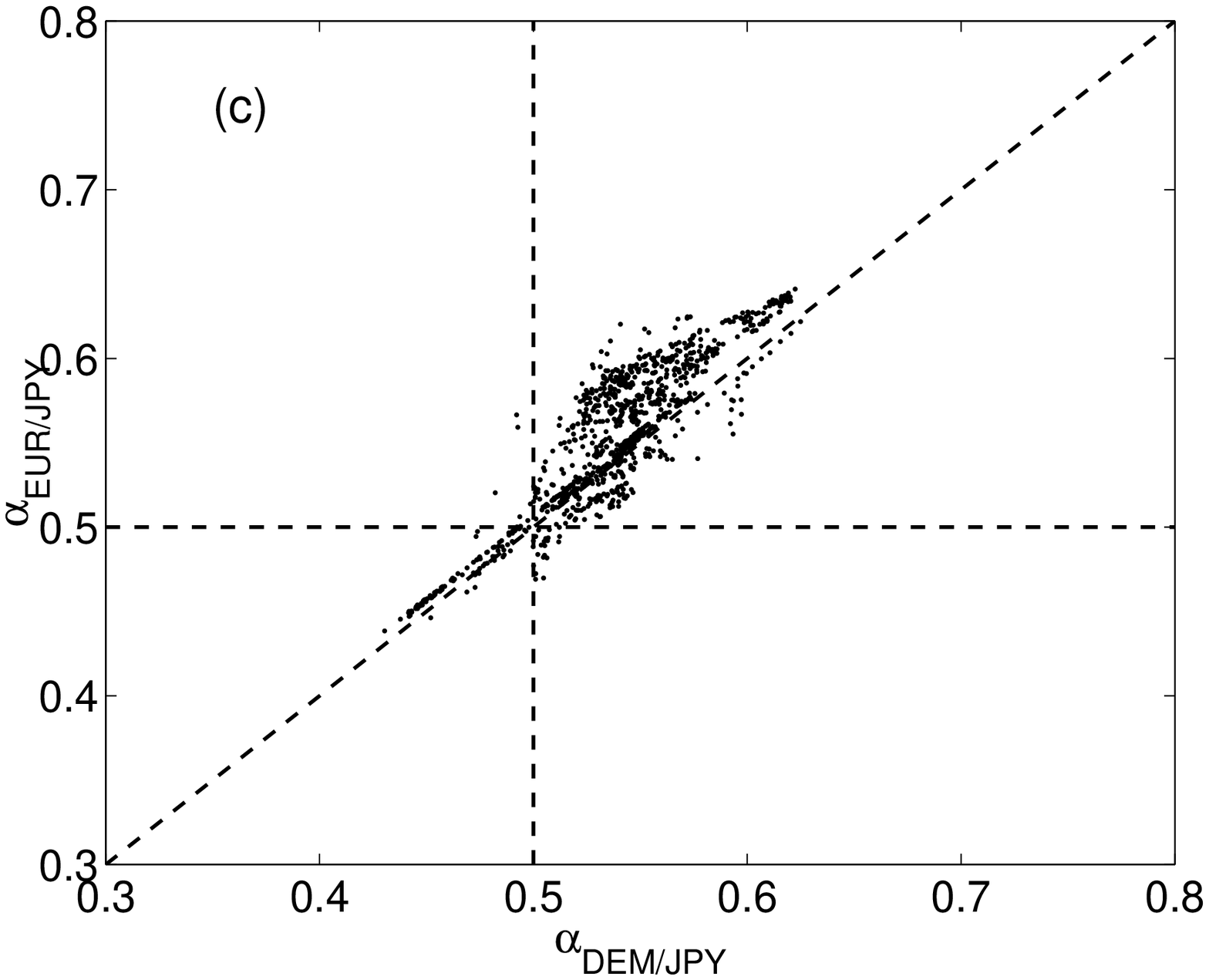} \hfill
\includegraphics[width=.36\textwidth]{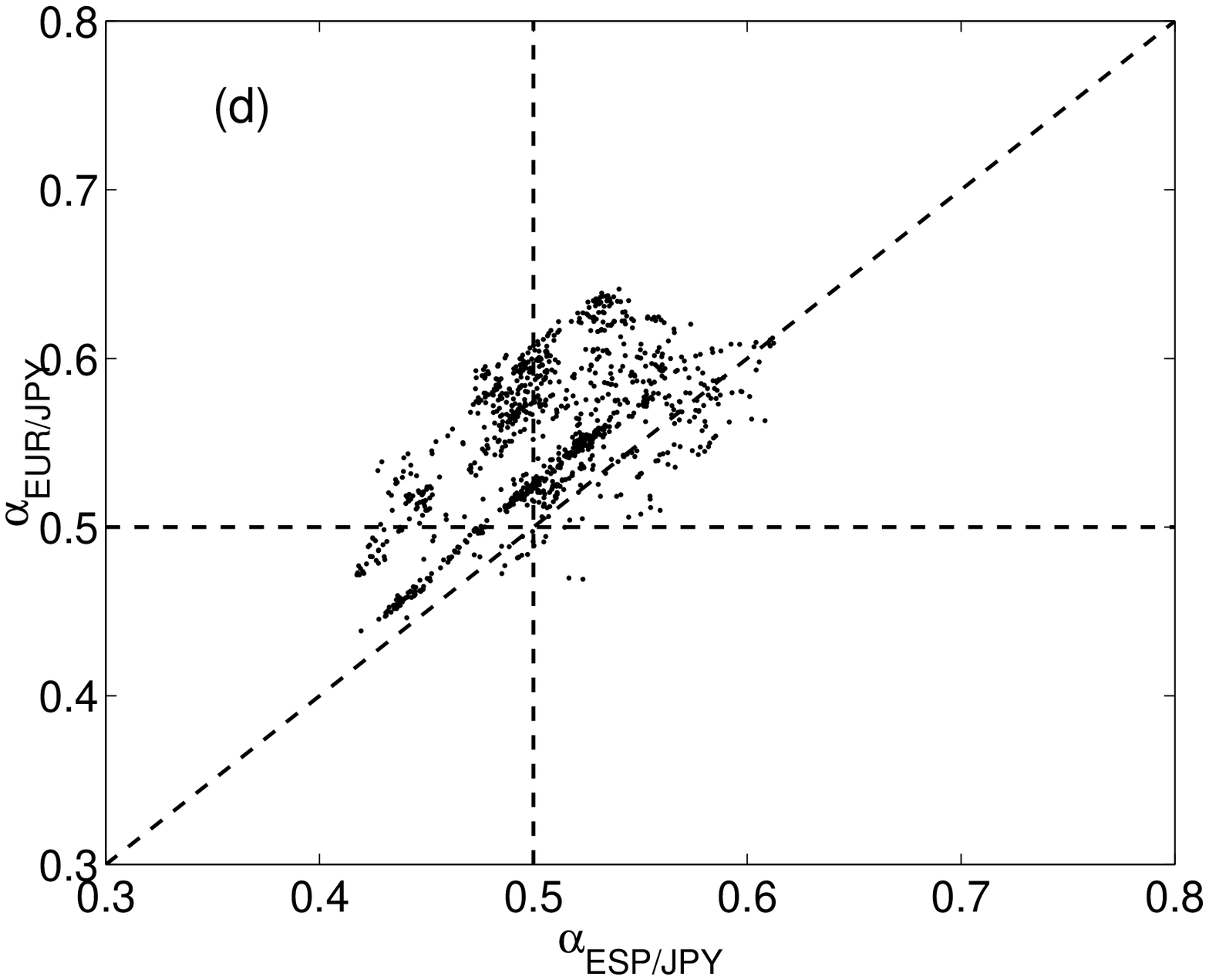} \vfill
\includegraphics[width=.36\textwidth]{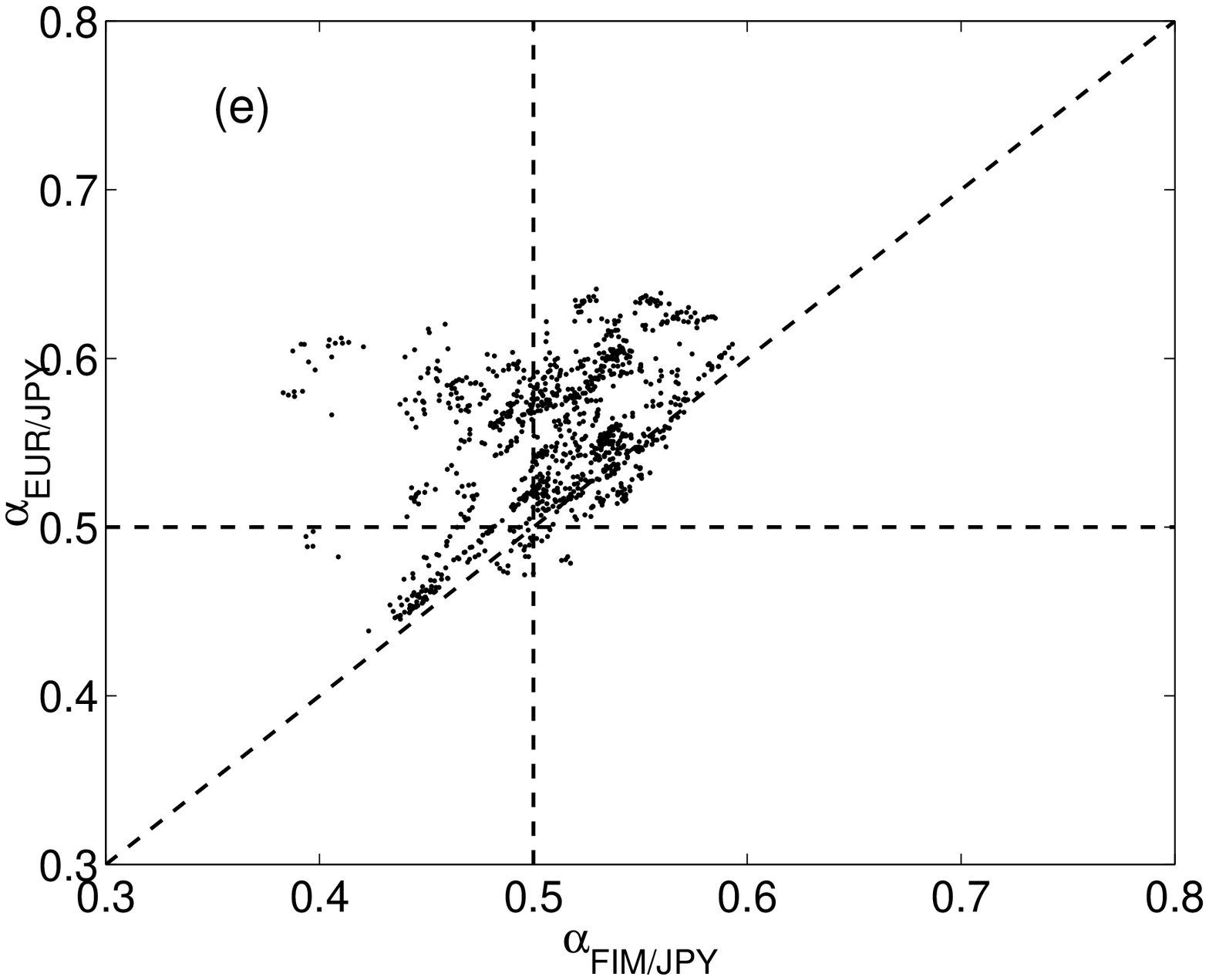} \hfill
\includegraphics[width=.36\textwidth]{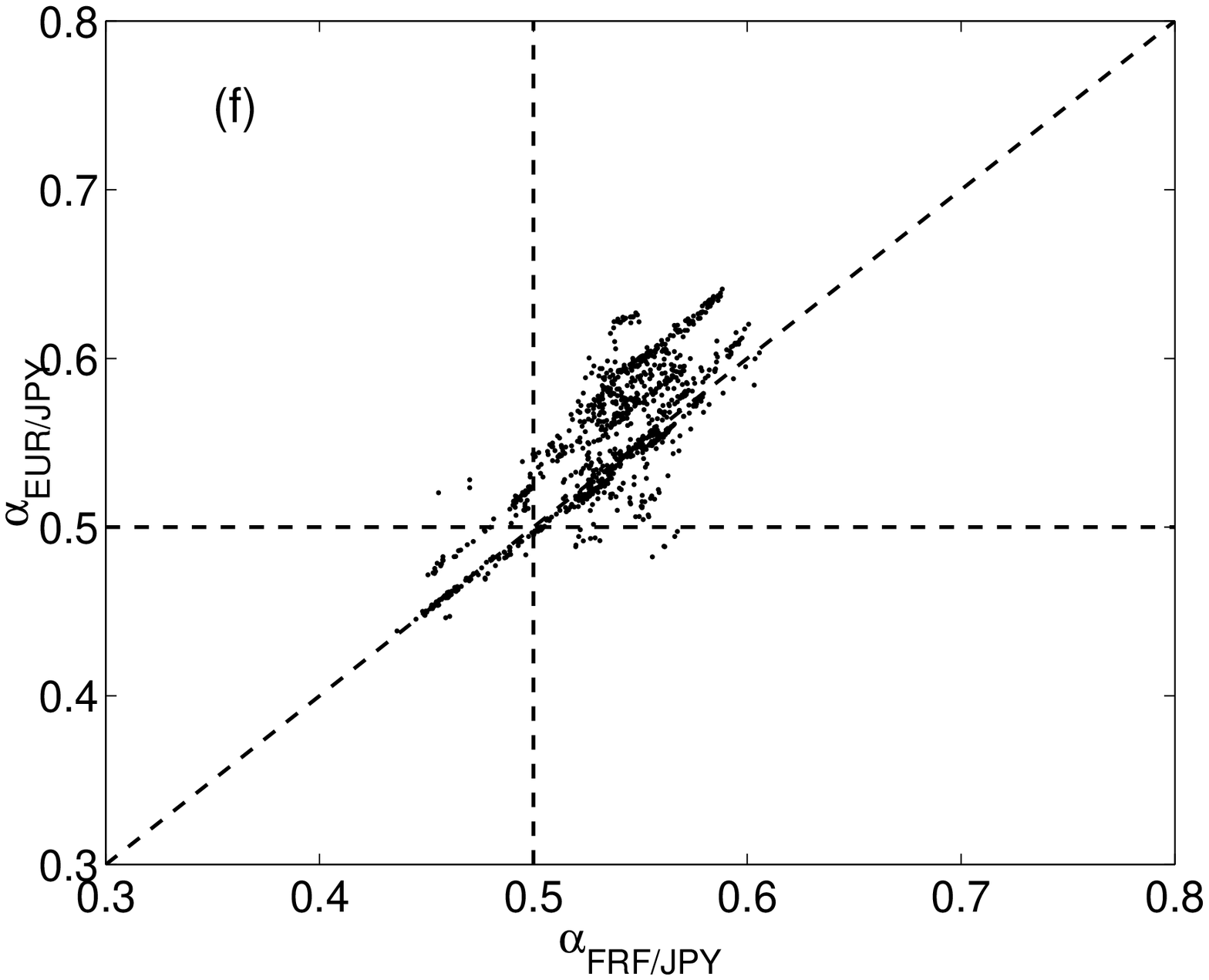} \vfill
\includegraphics[width=.36\textwidth]{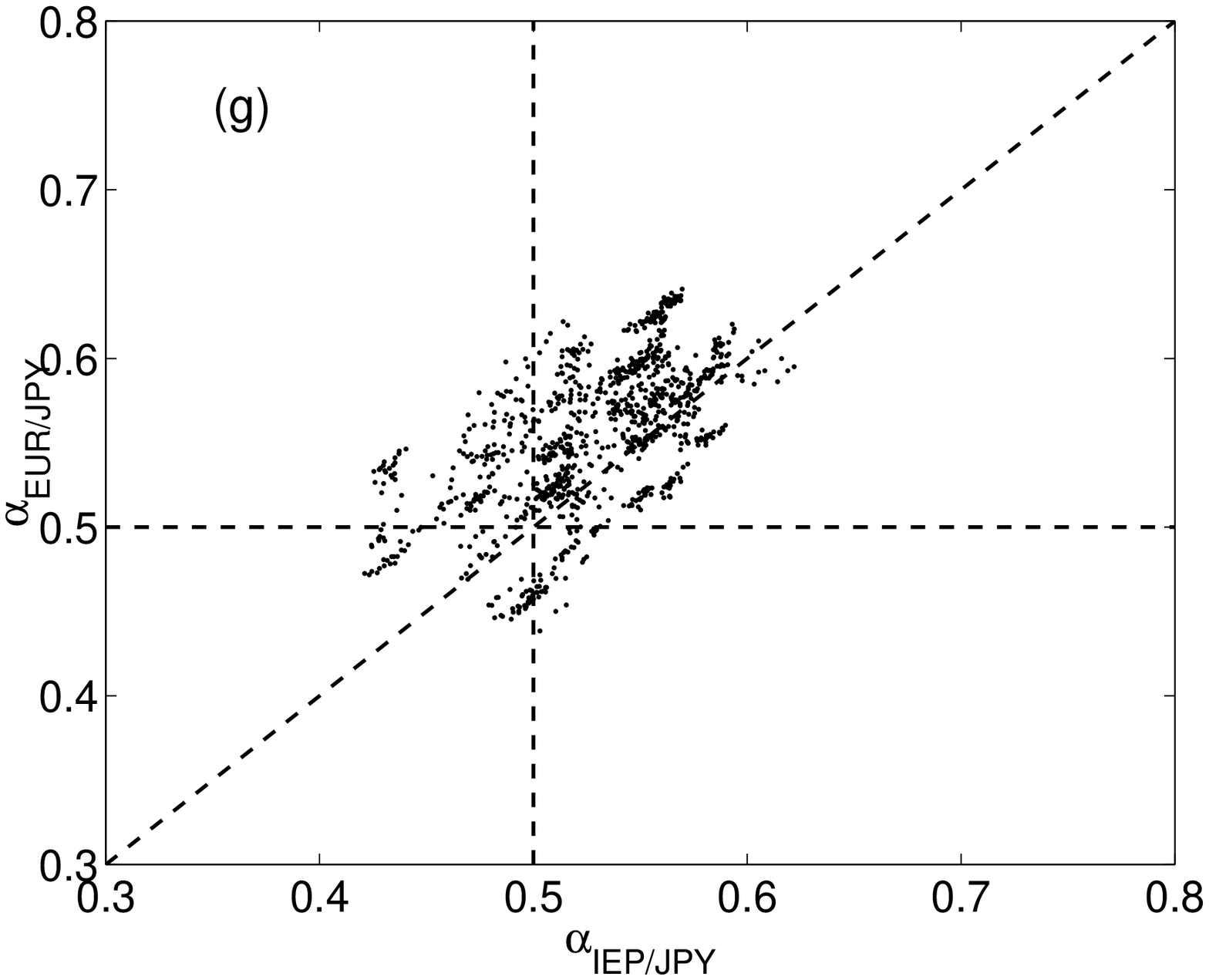} \hfill
\includegraphics[width=.36\textwidth]{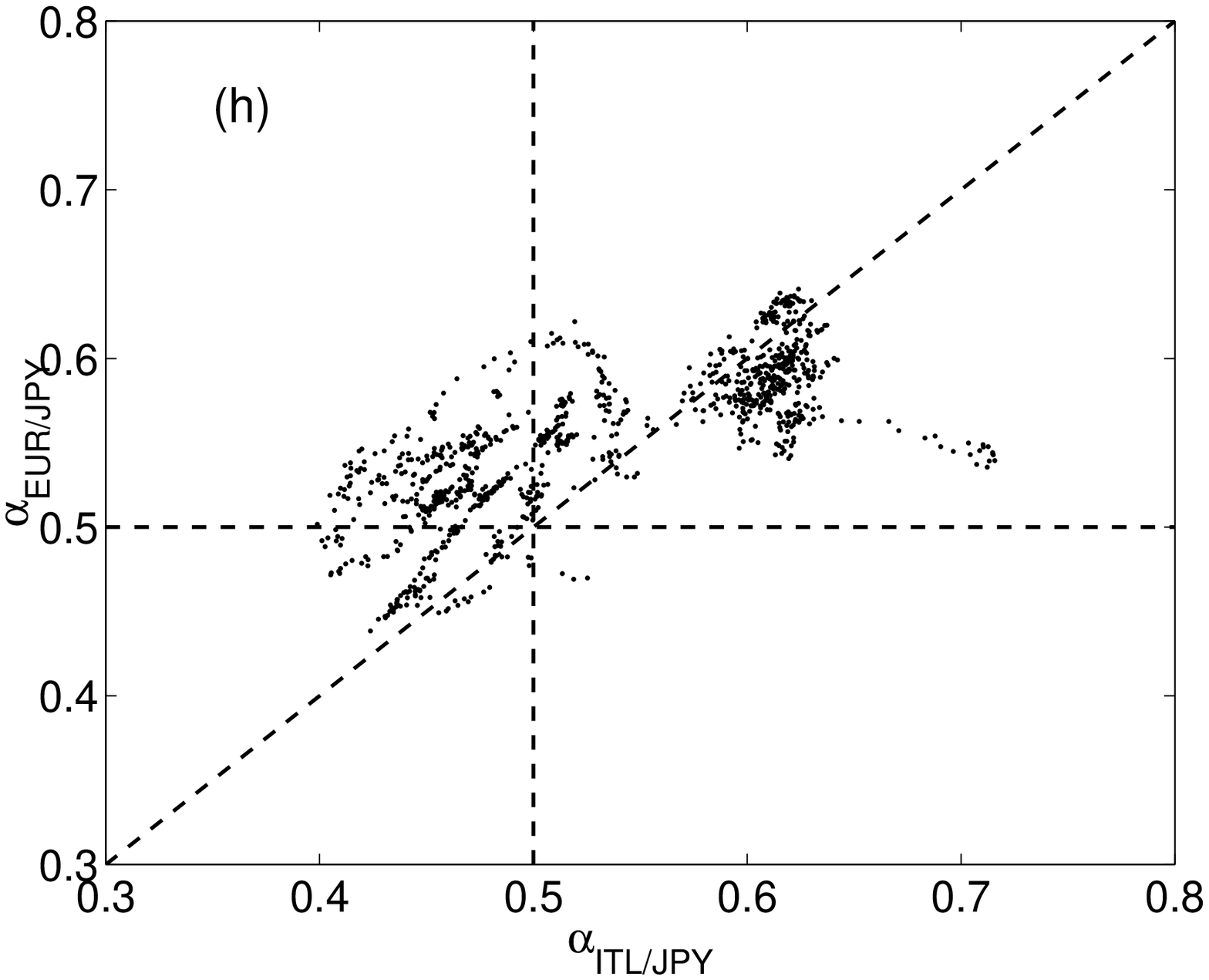} \vfill
\includegraphics[width=.36\textwidth]{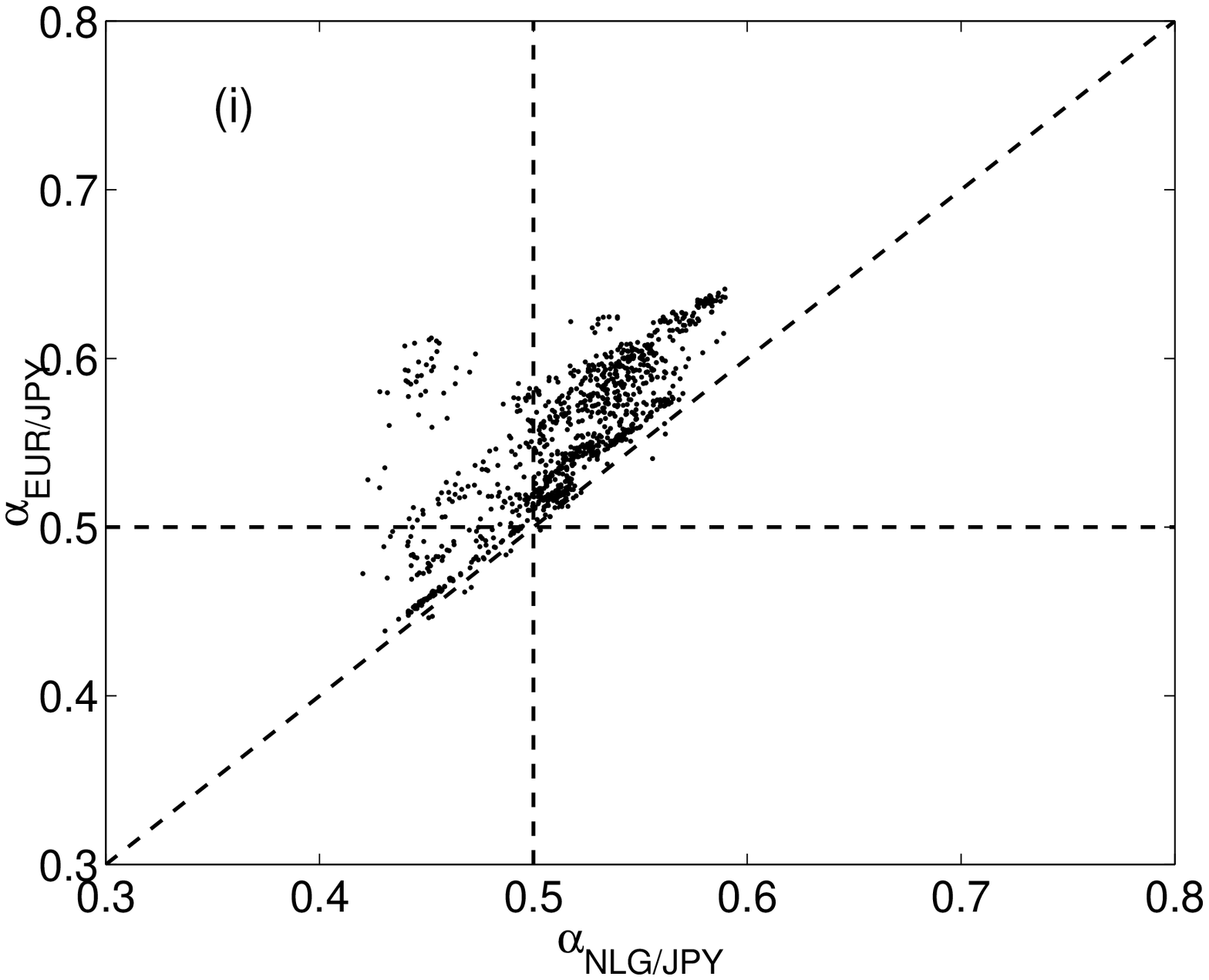} \hfill
\includegraphics[width=.36\textwidth]{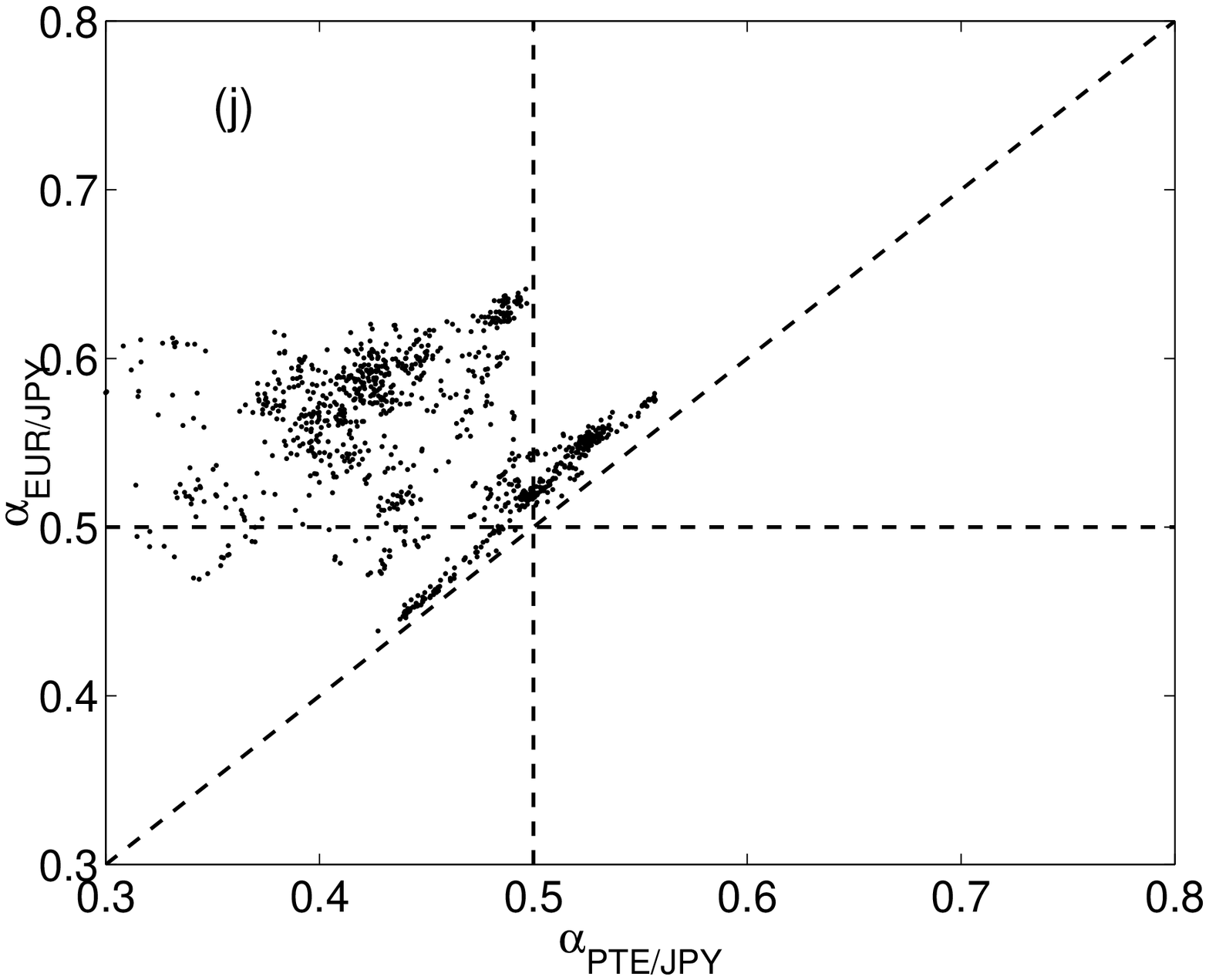} \caption{Structural 
correlation
diagram of the $DFA$ - $\alpha$ exponent for the exchange rate 
$EUR/JPY$ and the
$DFA$ - $\alpha$ exponent for the exchange rate with respect to $JPY$ of the 11
currencies forming one $EUR$; ($BEF$=$LUF$)} \label{eps4} \end{figure}

Looking for more diversified answers to our question on whether e.g. 
$DEM$ truly
controls the market, we have constructed a graphical correlation matrix of the
time-dependent $\alpha$ exponent for the various exchange rates of interest. In
Fig. 4(a--j), the correlation matrix is displayed for the time interval between
Jan. 1, 1993 and Jan. 1, 1999 for $\alpha_{EUR/JPY}$ {\it vs.}
$\alpha_{C_i/JPY}$, where $C_i$ stands for the 11 currencies that 
form the $EUR$;
($BEF$=$LUF$). As described elsewhere, e.g. in \cite{Ref2EUR}, such a 
diagram can
be divided into sectors through a horizontal, a vertical and perpendicular
diagonal lines crossing at (0.5,0.5). If the correlation is strong the cloud of
points should fall along the slope $=1$ line. This is clearly the case of the
relationship between $\alpha_{EUR/JPY}$ and $\alpha_{DEM/JPY}$, while 
the largest
spread is readily seen for $\alpha_{EUR/JPY}$ $vs.$ $\alpha_{PTE/JPY}$.

In order to assess additional features of the time dependent $\alpha$-exponents
of the exchange rate fluctuation correlations we have time averaged 
$\alpha$ for
the $EUR$ exchange rate with respect to $JPY$, and for each of the 11 eleven
currencies which form $EUR$, over the time interval [Jan. 1, 1993 - Jan. 1,
1999]. We present the results of $\alpha_{mean}$, $\alpha_{median}$,
$\alpha_{mean}/\alpha_{median}$ and the standard deviation $\sigma(\alpha)$ in
Table I. The {\it time evolution} of such quantities is given in Figs. 5(a--d),
where $\alpha_{mean}-\alpha_{median}$ rather than 
$\alpha_{mean}/\alpha_{median}$
is displayed for readability.

\begin{figure} \centering \includegraphics[width=.48\textwidth]{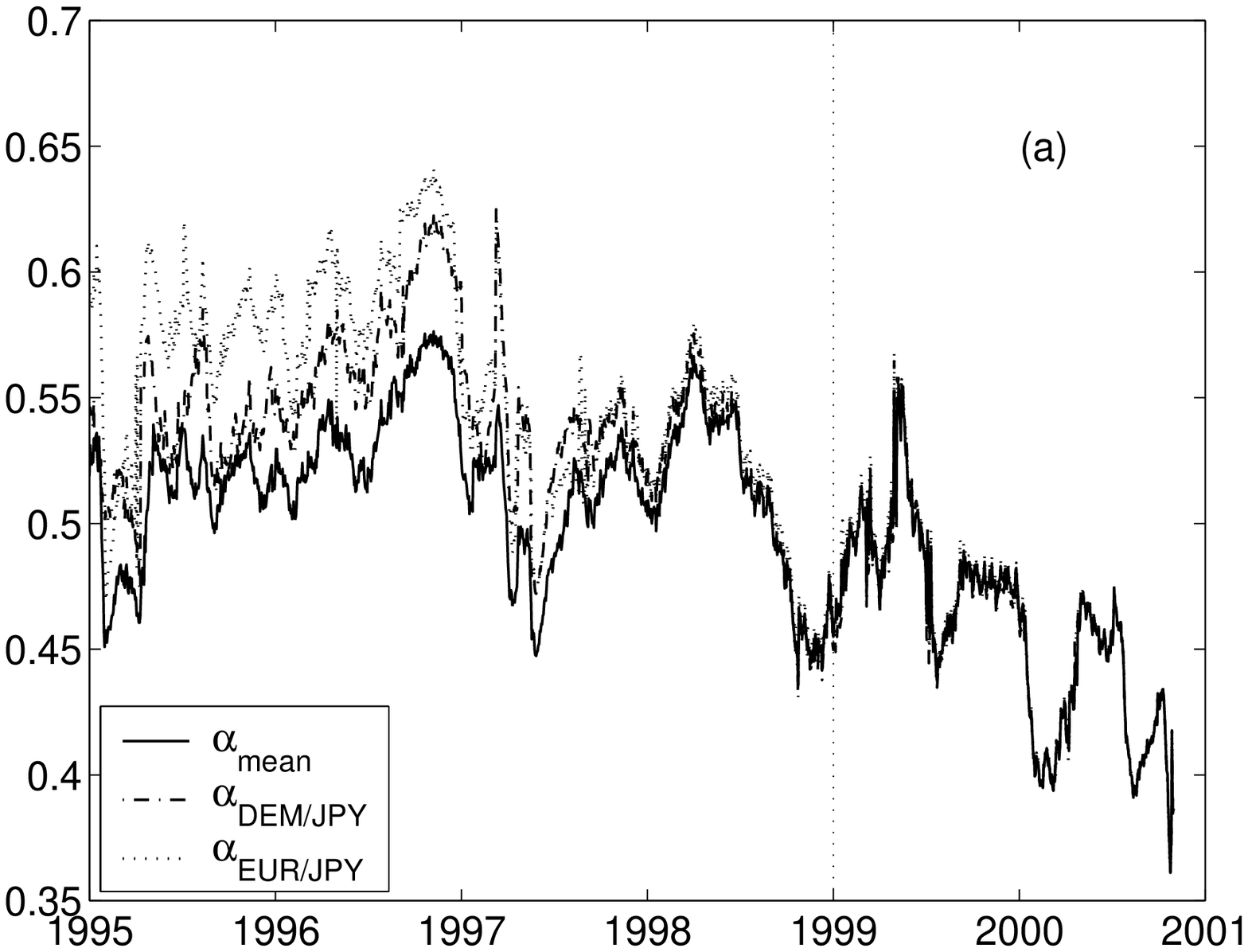}
\hfill \includegraphics[width=.48\textwidth]{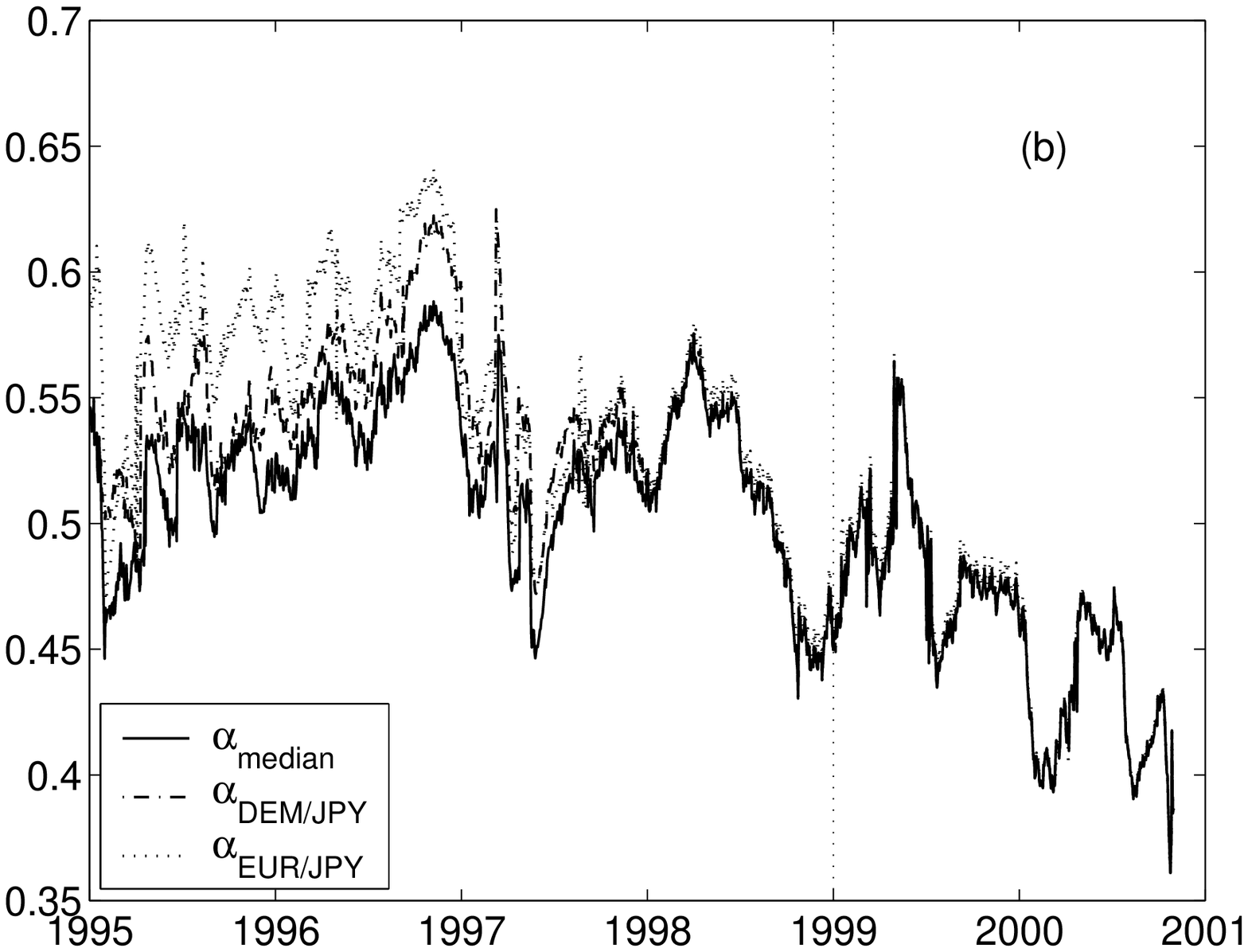} \vfill
\includegraphics[width=.48\textwidth]{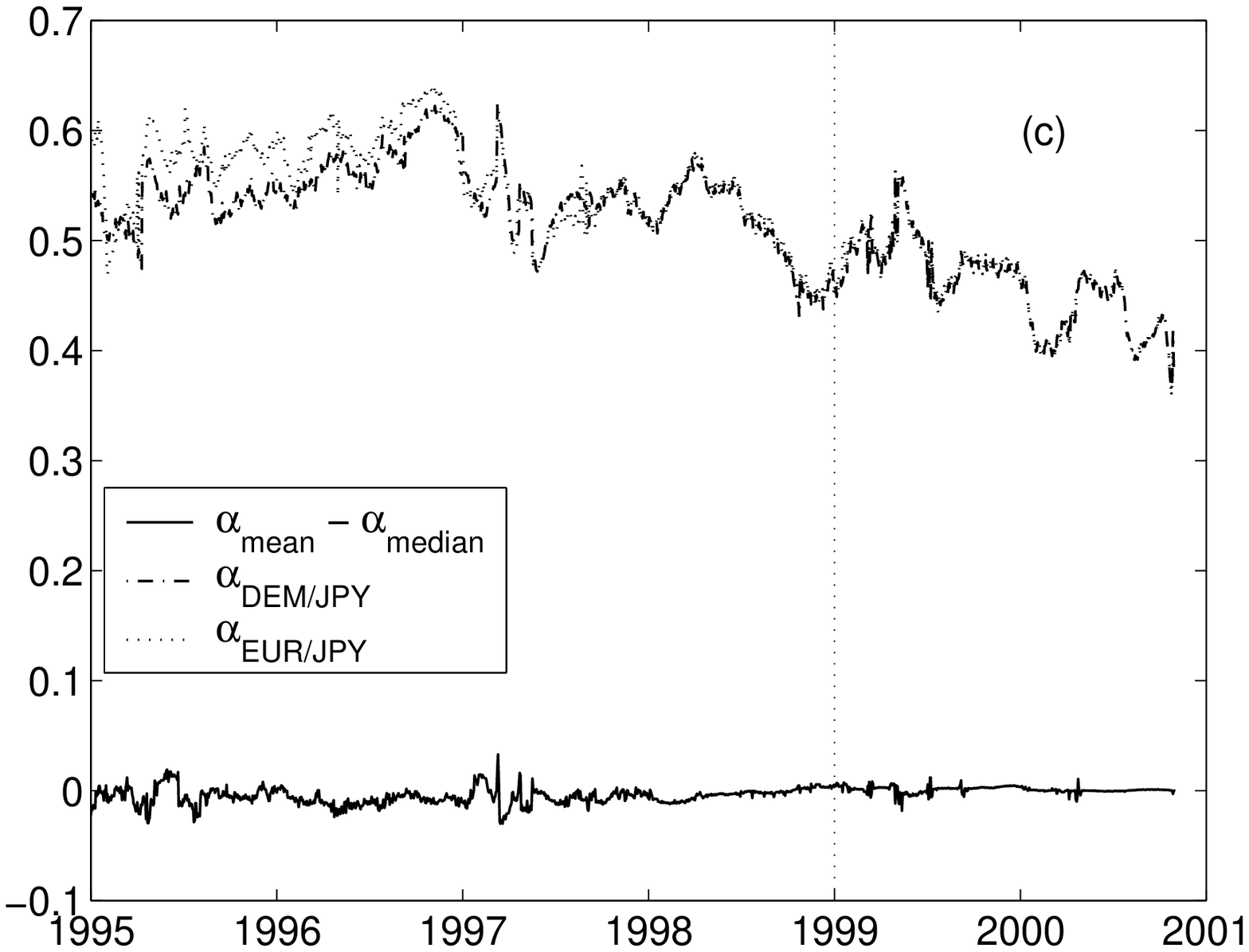} \hfill
\includegraphics[width=.48\textwidth]{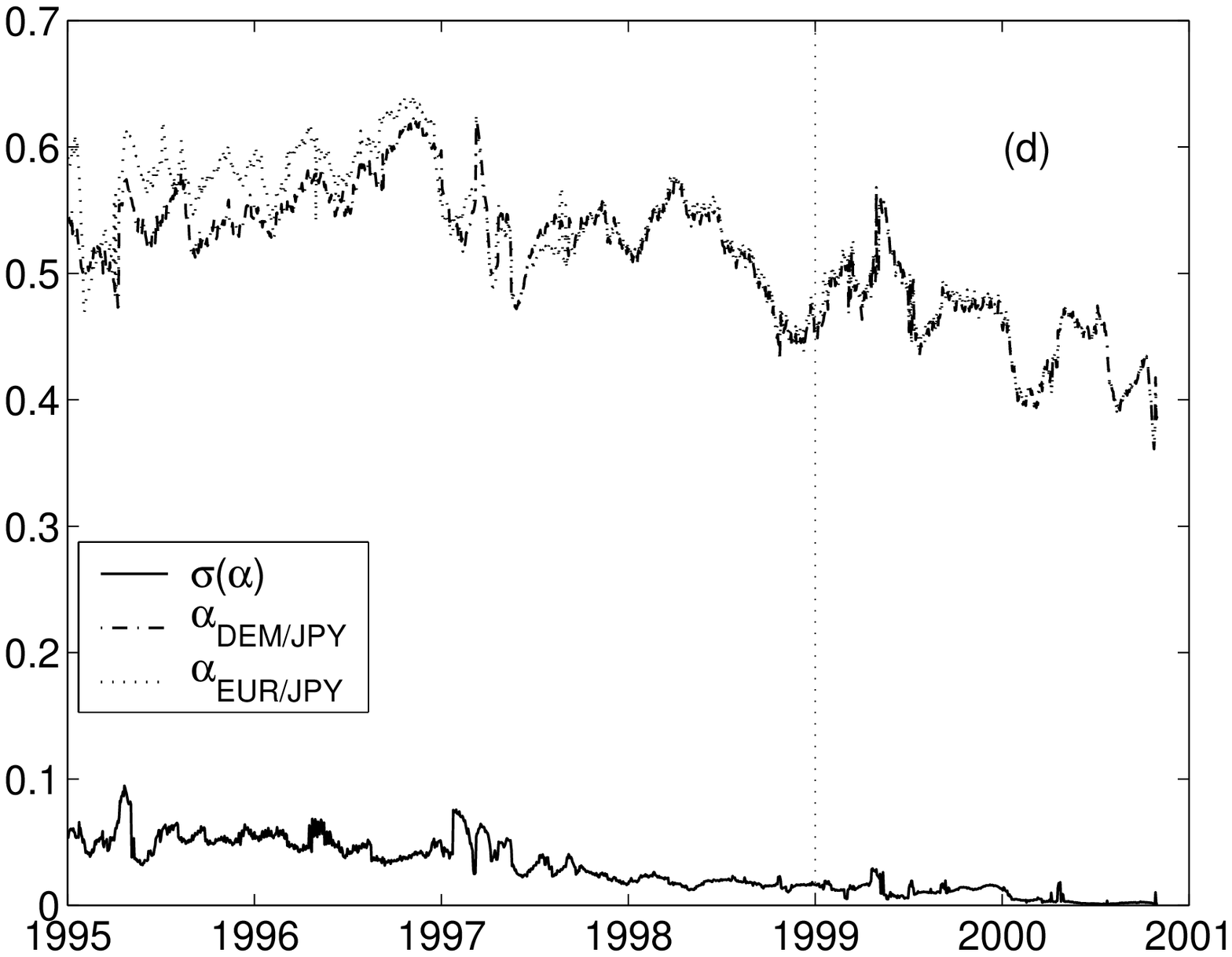} \caption{Time dependent mean,
median, mean-median and standard deviation of the $\alpha$-exponents resulting
from the time dependent $\alpha$ distributions of the eleven 
currencies that form
the $EUR$ and the corresponding time dependent quantities of the
$\alpha$-exponents for $EUR/JPY$ and $DEM/JPY$} \label{eps5} \end{figure}

Several remarks follow from the correlations shown with the structural diagrams
and the relations between the $\alpha_{mean}$ and $\alpha_{median}$. While the
structural diagrams for $ATS$, $PTE$, and $ITL$ with respect to $EUR$ show weak
or no correlation at all, the $\alpha_{mean}$ is equal within the error bars to
the $\alpha_{median}$ for $PTE$ and $ATS$ but $\alpha_{mean}$  and
$\alpha_{median}$ are markedly different from each other for $ITL$. 
On the other
hand, $\alpha_{mean}$ is not equal to $\alpha_{median}$ for $DEM$ but the
structural diagram shows very strong correlations between the time dependent
$\alpha$-exponents. It is clear that the leading currencies from the point of
view of the exchange rate fluctuations were $DEM$, and to a lesser 
extent $BEF$,
$FRF$, $IEP$, and $NLG$ while $PTE$ is far away from the main stream, i.e.
$\alpha \simeq 0.45$. We stress the quite small and quite large values of
$\sigma(\alpha)$ for $ATS$ and $ITL$ respectively, both having the largest
$\alpha_{mean}/\alpha_{median}$ ratio, both greater than unity in fact,
indicating e.g. specific national bank financial policies.

Let it be pointed out that $\alpha_{mean}$ follows $\alpha_{EUR/JPY}$ while
$\alpha_{median}$ follows the fluctuations of $\alpha_{DEM/JPY}$.

\begin{table}[ht] \centering \caption{Averaged values for the time 
interval [Jan. 1,
1993 ; Jan. 1, 1999] of the time dependent $\alpha$-exponents of the exchange
rate with respect to the Japanese Yen ($JPY$) for $EUR$ and those 11 currencies
forming the $EUR$ on Jan. 01, 99.} \renewcommand{\arraystretch}{1.4}
\setlength\tabcolsep{5pt} \begin{tabular}{ccccc} \hline 
Currency&$\alpha_{mean}$
& $\alpha_{median}$&$\alpha_{mean}$/$\alpha_{median}$ 
&$\sigma(\alpha)$ \\ \hline
EUR& 0.5541 & 0.5565 & 0.9956 & 0.0441 \\ \hline ATS& 0.4994 & 0.4971 
& 1.0046 &
0.0355 \\ BEF& 0.5264 & 0.5284 & 0.9961 & 0.0357 \\ DEM& 0.5389 & 
0.5408 & 0.9966
& 0.0361 \\ ESP& 0.5076 & 0.5062 & 1.0028 & 0.0398 \\ FIM& 0.5096 & 0.5136 &
0.9922 & 0.0388 \\ FRF& 0.5366 & 0.5408 & 0.9923 & 0.0325 \\ IEP& 
0.5282 & 0.5339
& 0.9893 & 0.0395 \\ ITL& 0.5348 & 0.5174 & 1.0337 & 0.0763 \\ LUF& 0.5264 &
0.5284 & 0.9961 & 0.0357 \\ NLG& 0.5168 & 0.5200 & 0.9937 & 0.0364 \\ 
PTE& 0.4483
& 0.4440 & 1.0097 & 0.0534 \\ \hline \end{tabular} \label{Tab1} \end{table}

\section{Conclusion}

We have thus studied a few aspects of the $EUR$ exchange rates from 
the point of
view of the fluctuations of the $EUR$ and the 11 currencies forming 
the $EUR$. We
have examined here the exchange rates toward $DKK$, $CHF$, $JPY$ and $USD$. The
central part of the distribution of the fluctuations can be fitted by 
a Gaussian,
while the distribution of the large fluctuations follows a power law. We have
observed that the $DEM$ is the strongest currency that has dominated the
correlations of the fluctuations in $EUR$ exchange rates with respect to $JPY$,
while $PTE$ was the most extreme one in the other direction.

\vskip 0.6cm

{\noindent \large \bf Acknowledgements} \vskip 0.6cm We are very 
grateful to the
organizers of the Symposium for their invitation. We gratefully thank the
Symposium sponsors for financial support.


\clearpage \addcontentsline{toc}{section}{Index} \flushbottom \printindex


\vskip 0.5cm \begin{thebibliography}{10} %
\addcontentsline{toc}{section}{References}

\bibitem{quoteEUR} http://pacific.commerce.ubc.ca/xr/euro/

\bibitem{rateEUR} http://pacific.commerce.ubc.ca/xr/euro/euro.html\#Rates

\bibitem{Ref1EUR} Ausloos~M., Ivanova~K. (2000) Introducing False 
$EUR$ and false
$EUR$ exchange rates. Physica A 286:353

\bibitem{DFA} Peng~C.-K., Buldyrev~S.V., Havlin~S., Simmons~M., Stanley~H.E.,
Goldberger~A.L. (1994) On the mosaic organization of DNA sequences. Phys Rev E
49:1685

\bibitem{kimaijmpc} Ausloos~M., Ivanova~K. (2001) Correlations Between
Reconstructed $EUR$ Exchange Rates vs. $CHF$, $DKK$, $GBP$, $JPY$ and 
$USD$. Int
J Mod Phys C (in press)

\bibitem{friedrich} Friedrich~R., Peincke~J., Renner~Ch. (2000) How to quantify
deterministic and random influences on the statistics of the foreign exchange
market. Phys Rev Lett 84:5224

\bibitem{gencay} Gencay~R., Selcuk~F., Whitcher~B. (2001) Scaling properties of
foreign exchange volatility. Physica A 289:249

\bibitem{castiglione} Castiglione~F., Pandey~R.~B., Stauffer~D. 
(2001) Effect of
trading momentum and price resistance on stock market dynamics: a Glauber Monte
Carlo simulation. Physica A 289:223

\bibitem{nvma} Vandewalle~N., Ausloos~M. (1997) Coherent and random 
sequences in
financial fluctuations. Physica A 246:454

\bibitem{Ref2EUR} Ausloos~M., Ivanova~K. (2001) False Euro ($FEUR$) 
exchange rate
correlated behaviors and investment strategy. Eur Phys J B (in press)

\end{thebibliography}
\end{document}